\documentclass[useAMS,usenatbib]{mn2e}

\usepackage{graphicx}
\usepackage{epsfig}
\usepackage{longtable}
\usepackage{times}

\usepackage{amsmath}
\usepackage{amssymb}
\def\XMM{{\em XMM--Newton}}

\def\RXTE{{\em RXTE}}

\def\Swift{{\em Swift}}
\def\ASCA{{\em ASCA}}

\def\pn{{\em pn}}

\def\pn{{\em pn}}

\def\IGR{IGR J01572-7259}
\def\rxj{RX J0059.2-7138}
\def\smc{SMC X-2}
\def\approxgt{\mathrel{\hbox{\rlap{\lower.55ex \hbox {$\sim$}}
        \kern-.3em \raise.4ex \hbox{$>$}}}}
\def\approxlt{\mathrel{\hbox{\rlap{\lower.55ex \hbox {$\sim$}}
        \kern-.3em \raise.4ex \hbox{$<$}}}}
\def\pdot {\dot P}

\def\flux {\mbox{erg cm$^{-2}$ s$^{-1}$}}
\def\lum {\mbox{erg s$^{-1}$}}
\def\nh{$N_{\rm H}$}
\def\ltsima{$\; \buildrel < \over \sim \;$}
\def\lsim{\lower.5ex\hbox{\ltsima}}
\def\gtsima{$\; \buildrel > \over \sim \;$}
\def\gsim{\lower.5ex\hbox{\gtsima}}
\def\msole{~M_{\odot}}

\def\countsec{\hbox{counts s$^{-1}$}}
\def\fph {ph cm$^{-2}$ s$^{-1}$}
\def\hcm {\hbox {\ifmmode $ atom cm$^{-2}\else atom cm$^{-2}$\fi}}

\def\chisqnu {$\chi^{2}_{\nu}$}

\newcommand{\be}{\begin{equation}}

\newcommand{\ee}{\end{equation}}

\title[Spectral analysis of \IGR]{Spectral analysis of \IGR\ during its 2016 outburst}
\vspace{-0.5 cm}
\author[La Palombara et al.]{N.~La Palombara$^{1}$\thanks{E-mail: nicola@iasf-milano.inaf.it}, P.~Esposito$^{2}$, S. Mereghetti$^{1}$, F.~Pintore$^{1}$, L.~Sidoli$^{1}$, and A.~Tiengo$^{1,3,4}$ \\
$^{1}$INAF, Istituto di Astrofisica Spaziale e Fisica Cosmica, via E.\ Bassini 15,   I-20133 Milano,  Italy   \\
$^{2}$Anton Pannekoek Institute for Astronomy, University of Amsterdam, Postbus 94249, NL-1090-GE Amsterdam, The Netherlands \\
$^{3}$IUSS, Istituto Universitario di Studi Superiori, piazza della Vittoria 15,  I-27100 Pavia, Italy \\
$^{4}$INFN, Sezione di Pavia, via A. Bassi 6, I-27100 Pavia, Italy \\
}

\begin{document}
\vspace{-0.5 cm}

\date{Accepted 2017 December 19. Received 2017 November 29; in original form 2017 October 16}
\vspace{-0.5 cm}

\pagerange{\pageref{firstpage}--\pageref{lastpage}} \pubyear{2017}
\vspace{-0.5 cm}

\maketitle
\vspace{-0.5 cm}

\label{firstpage}
\vspace{-0.5 cm}

\begin{abstract}
We report on the results of the \XMM\ observation of \IGR\ during its most recent outburst in 2016 May, the first since 2008. The source reached a flux $f \sim 10^{-10}$ \flux, which allowed us to perform a detailed analysis of its timing and spectral properties. We obtained a pulse period $P_{\rm spin}$ = 11.58208(2) s. The pulse profile is double peaked and strongly energy dependent, as the second peak is prominent only at low energies and the pulsed fraction increases with energy. The main spectral component is a power-law model, but at low energies we also detected a soft thermal component, which can be described with either a blackbody or a hot plasma model. Both the EPIC and RGS spectra show several emission lines, which can be identified with the transition lines of ionized N, O, Ne, and Fe and cannot be described with a thermal emission model. The phase-resolved spectral analysis showed that the flux of both the soft excess and the emission lines vary with the pulse phase: the soft excess disappears in the first pulse and becomes significant only in the second, where also the Fe line is stronger. This variability is difficult to explain with emission from a hot plasma, while the reprocessing of the primary X-ray emission at the inner edge of the accretion disk provides a realiable scenario. On the other hand, the narrow emission lines can be due to the presence of photoionized matter around the accreting source.
\end{abstract}

\vspace{-0.5 cm}

        \section{Introduction\label{intro}}

\IGR\ (SXP 11.6) is a poorly studied transient X-ray source which was discovered with INTEGRAL in 2008 in the Magellanic Bridge \citep{Coe+08}. It was detected with both the IBIS and JEM-X instruments, with a maximum flux $f_{\rm (20-60~keV)}$ =  (3.3$\pm$0.3)$\times10^{-11}$ erg cm$^{-2}$ s$^{-1}$ and $f_{\rm (3-10~keV)}$ = (1.6$\pm$0.5)$\times10^{-12}$ erg cm$^{-2}$ s$^{-1}$, respectively. Follow-up observations performed with \Swift\ and \RXTE\ led to a precise source localization and to the detection of strong pulsations with $P_{\rm spin}$ = 11.57809 $\pm$ 0.00002 s. A single star (USNO-B1 0170-0064697, \citealt{Monet+03}), with B and R magnitudes of 15.48 and 15.51, respectively, was found within the \Swift/XRT error circle. The broadband spectrum obtained with the \Swift/XRT and the IBIS data was modelled with an exponentially cut-off power law, with photon index $\Gamma = 0.4\pm0.2$ and folding energy $E_{\rm f} = 8^{+5}_{-3}$ keV. All these results suggested that the source was a new Be/X-ray binary (BeXRB, \citealt{McBride+10}).

The timing analysis of 88 months of survey data collected with \Swift/BAT led to the discovery of periodic modulation of the hard-X-ray emission, with a period of 35.6 $\pm$ 0.5 d and a significance of 6.1 $\sigma$ \citep{Segreto+13}. The BAT light curve folded at this period showed a single, wide peak and a minimum which was consistent with zero intensity: this suggested that the accretion of matter onto the neutron star (NS) occurs for most of the orbit. The centroid of the light-curve minimum occurs at the orbital phase $\Phi_{\rm min} = 1.01 \pm 0.02$, corresponding to MJD = (55225.2 $\pm$ 0.7) $\pm~nP_{\rm orbit}$. If the periodicity of 35.6 d is attributed to the orbit of the binary system, the position of \IGR\ in the Corbet diagram $P_{\rm spin} - P_{\rm orbit}$ \citep{Corbet86} is well within the region of the BeXRBs. From the spectral point of view, the combined XRT+BAT spectrum was fit with an exponentially cut-off power law very similar to that used for the XRT+IBIS spectrum, with an upper limit \nh\ $< 5\times10^{20}$ cm$^{-2}$ on the hydrogen column density. Comparison with the total Galactic absorption in this direction, which is \nh\ = $6\times10^{20}$ cm$^{-2}$ according to \citet{Kalberla+05} or \nh\ $\simeq 3.5\times10^{20}$ cm$^{-2}$ according to \citet{DickeyLockman90}, suggests that there is no, or very low, intrinsic absorption in the source and/or locally in the SMC bridge.

The optical counterpart of \IGR\ was investigated in the I band with three seasons of OGLE-IV observations. \citet{Schmidtke+13} reported on the detection of modulations at $P_{\rm optical}$ = 35.1 $\pm$ 0.1 d in the OGLE light curve. Moreover, they observed that, using the ephemeris obtained by \citet{Segreto+13} with the BAT data, the X-ray and optical peaks were found to be nearly coincident, differing in phase by $<$ 0.1 optical cycles. This reinforced the interpretation that the periodicity of 35 d is an orbital signature.


After the \Swift/XRT observations carried out in October 2010 \citep{Segreto+13}, no further observations of \IGR\ were performed until 2016 April 15 (MJD = 57493), when \Swift/BAT detected an outburst from this source \citep{Krimm+16}. Afterwards, the source flux increased from $\sim 1.4 \times 10^{-10}$ erg cm$^{-2}$ s$^{-1}$ up to the flux peak of $\sim 2.4 \times 10^{-10}$ erg cm$^{-2}$ s$^{-1}$ on 2016 April 25 (MJD = 57503). This prompted us to trigger our \XMM\ Target--of--Opportunity programme for the observation of bright transient pulsars in the Small Magellanic Cloud (SMC), thus obtaining an observation of \IGR\ in 2016 May. Here we report on the results obtained. 

 	 \section{Observation and Data Reduction}
         \label{data}

\IGR\ was observed with \XMM\ on 2016 May 7 (MJD = 57515), when it was at the beginning of its outburst. The total exposure time was 28 ks and the three EPIC focal-plane cameras (one for each telescope), i.e.~one \pn\ \citep{Struder+01} and two MOS \citep{Turner+01}, were all operated in \textit{Small Window} mode. The time resolution was 5.7 ms for the \pn\ camera \citep{Struder+01} and 0.3 s for the two MOS cameras \citep{Turner+01}. For all cameras the Thin filter was used. The Reflection Grating Spectrometer (RGS) was operated in \textit{Spectroscopy} mode \citep{denHerder+01}.

We used version 15 of the \XMM~{\em Science Analysis System} (\textsc{sas}) to process the event files. After the standard pipeline processing, we searched for possible intervals of high instrumental background. We found that the last $\simeq$ 7 ks of the observation were characterized by a high background level. For the spectral analysis we rejected the data collected during this time range, since it could affect our results. Moreover, the net exposure time was further reduced by an outage of the ground station after the first 20 ks of the observation: this produced a time gap of 1.79 ks in all the instruments, during which the data were lost. Taking into account the rejected data, the time gap and the instrumental dead time (29 \% and 2.5 \% for \pn\ and MOS, respectively), the effective exposure time was $\simeq$ 15 ks for the \pn\ camera and $\simeq$ 21 ks for the MOS and the RGS instruments. In Table~\ref{observation} we summarise of the \XMM\ observation.

For the analysis of the EPIC data, we selected events with pattern in the range 0--4 (mono-- and bi--pixel events) for the \pn\ camera and 0--12 (from 1-- to 4--pixel events) for the two MOS. We considered circular extraction regions around the source position with a radius of 30, 40, and 50 arcsec for the \pn, MOS1, and MOS2 camera, respectively. In all cases the radius size was limited by the CCD edges or dark columns. Although the source count rate (CR) was very high (Table~\ref{observation}), we checked that neither the \pn\ nor the MOS data were affected by photon pile-up. To this aim, for each camera we compared spectra obtained with different pattern selections (only mono- or bi-pixel events) and with or wihout the excision of the core of the Point Spread Function (PSF), where the possible pile-up is higher. In all cases we found no differences among the various spectra. Moreover, we performed a fit of the radial profile with a King function, and found no significant residuals. Therefore, we concluded that the pile-up was negligible and selected events from the whole circular region. For each camera, background events were selected from circular regions offset from the target position, and free of sources.

\begin{table*}
\caption{Summary of the \XMM\ observation (ID 0780312701) of \IGR.}\label{observation}
\begin{center}
\begin{tabular}{cccccc} \\ \hline
Instrument	& Filter	& Mode		& Net Exposure Time (ks)	& Extraction Radius (arcsec)	& Net Count Rate (\countsec) 	\\ \hline
\pn\		& Thin 1	& Small Window	& 14.8				& 30				& 14.34$\pm$0.03		\\
MOS1		& Thin 1	& Small Window	& 20.9				& 40				& 3.78$\pm$0.01			\\
MOS2		& Thin 1	& Small Window	& 20.8				& 50				& 4.41$\pm$0.01			\\
RGS1		& - 		& Spectroscopy	& 21.5				& -				& 0.274$\pm$0.003		\\
RGS2		& - 		& Spectroscopy	& 21.4				& -				& 0.090$\pm$0.002		\\ \hline
\end{tabular}
\end{center}
\end{table*}

 	 \section{Timing analysis}
         \label{timing}

\begin{figure}
\begin{center}
\includegraphics[width=5.5cm,angle=-90]{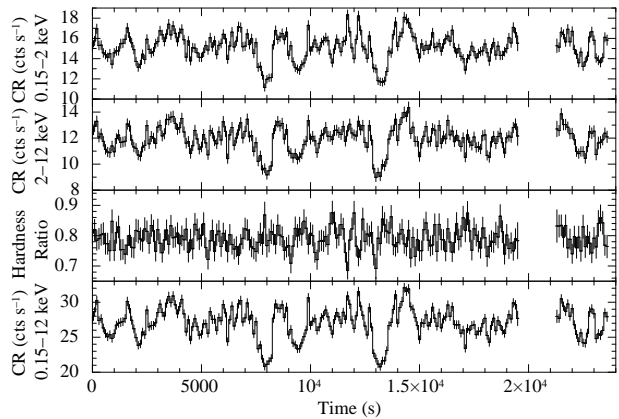}
\caption{Background-subtracted light curves of \IGR\ in the energy ranges 0.15-2, 2-12, and 0.15-12 keV, with a time bin of 100 s.}
\label{lc}
\end{center}
\vspace{-0.75 cm}
\end{figure}

For the timing analysis, we used the \textsc{sas} tool \textsc{barycenter} to report the EPIC event arrival times to the Solar system barycenter. To investigate the source flux and spectral variability over the whole \XMM\ observation, for each EPIC camera we accumulated three light curves (with a time binning of 100 s) in the 0.15-2 keV (soft), 2-12 keV (hard), and 0.15-12 (total) energy ranges. Then, we used the \textsc{sas} tool \textsc{epiclccorr} to correct each light curve for the background and the extraction region. In this way, we found that the average CR in the total range was $\simeq$ 4.8, 4.9, and 17.3 cts s$^{-1}$, for the MOS1, MOS2, and \pn\ cameras, respectively. In Fig.~\ref{lc} we report the cumulative light curves in the three energy ranges, obtained by summing the light curves of the individual cameras, together with the hardness-ratio (HR) of the hard (H) to soft (S) light curves (computed as H/S). The average CR in the soft and hard energy ranges was 15.1 and 11.9 cts s$^{-1}$, respectively. The figure shows that, in both ranges, the source was highly variable over short timescales, since there are CR variations of up to $\sim$ 20 \% between consecutive time bins. However, there is no evidence of long-term evolution, since the CR shows no incresing/decreasing trend along the observation. Also the HR shows large bin-to-bin variability, but without any long time-scale trend; moreover, there is no correlation between the HR and the CR.

\begin{figure}
\begin{center}
\includegraphics[width=5.5cm,angle=-90]{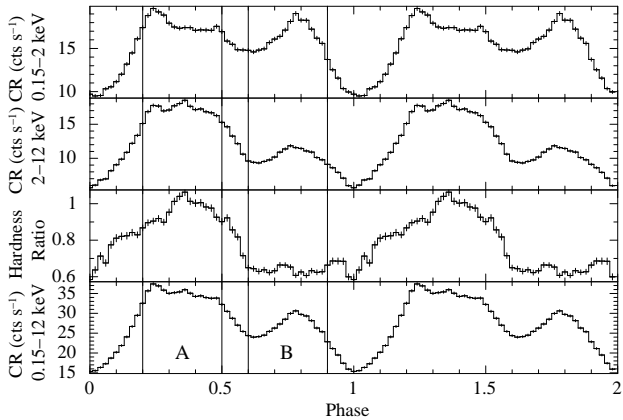}
\caption{Background-subtracted light curves of \IGR, in the energy ranges 0.15-2, 2-12, and 0.15-12 keV, folded at the best-fitting period $P$ = 11.58208 s. The vertical lines identify the two phase intervals (A and B) considered for the phase-resolved spectral analysis (section \ref{resolved_spectral_analysis}).}
\label{flc2E}
\end{center}
\vspace{-0.75 cm}
\end{figure}

To measure the pulse period of the source, the datasets of the barycenter-corrected events of the three instruments were combined together. Then, we used a standard phase-fitting technique and obtained a best-fitting period of $P$ = 11.58208(2) s. In Fig.~\ref{flc2E} we report the three light curves and the HR folded at the best-fitting period. In both the soft and hard energy ranges the pulse profile is rather smooth and shows two broad peaks, separated by an absolute minimum at $\Phi \simeq$ 1 and a secondary minimum at $\Phi \simeq$ 0.62. The profile is strongly energy dependent. In the soft range the two peaks have comparable amplitudes, while the second peak is much lower than the first in the hard range. Moreover, the structure of the pulse profile is more complex in the hard energy range: here, after the first peak corresponding to the highest peak in the soft energy range, there is also a second peak that is slightly brighter than the first. The HR increses with the CR along the first peak, while it is almost constant at its minimum value along the second peak. Also the average pulsed fraction, defined as PF = (CR$_{\rm max}$ - CR$_{\rm min}$)/(2$\times$CR$_{\rm average}$) \citep{Kohno+00}, is strongly energy dependent, since it is $\simeq$ 33 \% for the soft range and $\simeq$ 53 \% for the hard range.

\begin{figure}
\begin{center}
\includegraphics[width=5.5cm,angle=-90]{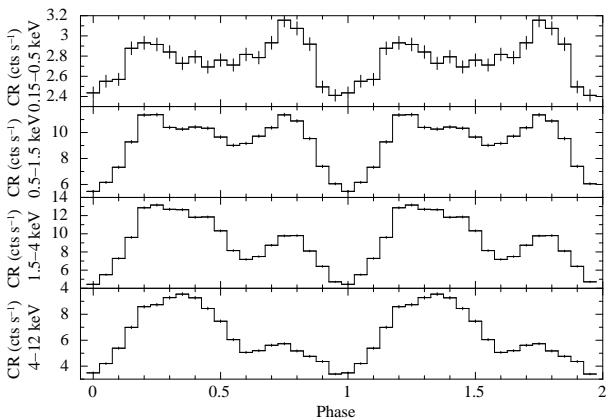}
\caption{Pulse profile of \IGR\ in the energy ranges 0.15-0.5, 0.5-1.5, 1.5-4.0, and 4.0-12.0 keV.}
\label{flc4E}
\end{center}
\vspace{-0.75 cm}
\end{figure}

The energy dependence of the pulse profile is better appreciable in Fig.~\ref{flc4E}, where we report the folded light curve in four, narrower energy bands. It clearly shows that the relative height of the second peak increases at decreasing energies, and that it overcomes the first peak for E $<$ 0.5 keV. On the other hand, the flux variability along the first peak strongly increases with energy: while the CR varies of only $\sim$ 20 \% at E $<$ 0.5 keV, it increases of a factor $\sim$ 3 for E $>$ 4 keV. Also the average PF varies with energy, increasing from 14 \% at E $<$ 0.5 keV up to 49 \% at E $>$ 4 keV.

\vspace{-0.5 cm}

  	\section{EPIC spectroscopy}
	\label{EPIC}

Since we found no strong evidence for intensity or spectral variability in \IGR\ on the observation time scale, for each EPIC camera we extracted a source time-averaged spectrum over the whole exposure. To this aim, we adopted the same extraction parameters used for the light curves, and each spectrum was rebinned so to obtain a significance of at least 3 $\sigma$ for each energy bin. The applicable response matrices and ancillary files were generated using the \textsc{sas} tasks \textsc{rmfgen} and \textsc{arfgen}, respectively. The spectral analysis was performed in the energy range 0.2-12 keV, using version 12.9.1 of \textsc{xspec}. In the following, all spectral uncertainties and upper limits are given at the 90 \% confidence level for one interesting parameter. Since \IGR\ is in the Magellanic Bridge, its distance has an intermediate value between the LMC distance of 50 kpc \citep{Pietrzynski+13} and the SMC distance of 62 kpc \citep{Graczyk+14}. We assumed a source distance of 56 kpc, although the most recent measurements of the structure of the SMC \citep{Scowcroft+16} imply that it might be an overestimate of the real value. After checking that separate fits of the three EPIC cameras gave consistent results, we fitted them simultaneously to improve the quality of the statistics. To this aim, we introduced free relative normalizations between the three cameras, to account for uncertainties in instrumental responses: the normalization factors for the MOS spectra relative to PN (factor fixed at 1) were 0.889$\pm$0.007 for MOS1 and 1.017$\pm$0.008 for MOS2. In the spectral fitting we adopted the interstellar abundances of \citet{WilmsAllenMcCray00} and the photoelectric absorption cross sections of \citet{Verner+96}, using the absorption model \textsc{tbnew} in \textsc{xspec}.

\begin{figure}
\includegraphics[height=8.5cm,angle=-90]{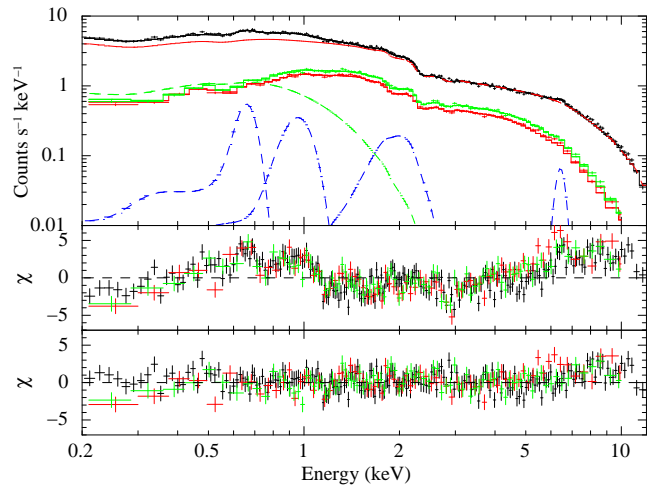}
\caption{Time-averaged spectrum of \IGR. \pn, MOS1, and MOS2 data are reported in black, red, and green, respectively. Upper panel: superposition of the EPIC spectra with (for the \pn\ spectrum only) the best-fitting \textsc{pl+bb} model (red and green dashed lines) plus the additional gaussian components (blue dashed lines). Middle panel: data-model residuals in the case of the fit with a simple \textsc{pl} model. Lower panel: data-model residuals in the case of the best-fitting model.}\label{epic_spectrum}
\end{figure}

It was not possible to obtain an acceptable description of the source spectrum with a single-component model. The spectral fit with an absorbed power-law (PL) provided a best-fit model with \chisqnu/d.o.f. = 1.32/3281, and revealed several spectral structures (Fig.~\ref{epic_spectrum}, medium panel): (1) a significant soft excess (SE) at the low-energy end of the spectrum; (2) a broad emission feature in the energy range 0.9-1.0 keV; (3) two narrow emission features at, respectively, $\sim$ 0.65 and $\sim$ 6.5 keV; (4) a narrow absorption feature at $\sim$ 2.9 keV. In addition, only the \pn\ spectrum showed an emission feature at $\sim$ 2 keV: it is very likely due to residual calibration uncertainties around the Au edge, and we modelled it with a Gaussian component \citep{DiazTrigo+14}. On the other hand, for both the emission features at $\sim$ 0.65 and 0.9-1.0 keV a calibration/instrumental origin can be excluded, since similar features were clearly also detected in the RGS spectra (see below).

It was possible to describe both the soft excess and the feature at 0.9-1.0 keV in two different ways: either with a blackbody (BB) component plus a Gaussian line or with an emission spectrum from collisionally ionized gas (\textsc{apec} model in \textsc{xspec}). In the first case, the fit of the overall spectrum required a PL+BB model to describe the spectral continuum, four additional Gaussian components to describe the emission features, and one multiplicative component (\textsc{gabs} in \textsc{xspec}) to describe the absorption feature (Fig.~\ref{epic_spectrum}, upper and lower panels). In the second case, the adoption of a single APEC component allowed us to account for both the soft excess and the emission feature at 0.9-1 keV. On the other hand, it revealed the presence of an additional very narrow emission feature at $\simeq$ 0.5 keV and of an additional absorption feature at $\simeq$ 1 keV. Therefore, the description of the spectral continuum with an absorbed PL+APEC model required four additional Gaussian components and two GABS components. 

In Table~\ref{epic_fit} we report the best-fitting parameters obtained for both spectral models. In the case of the APEC component, we considered two different options for the metal abundance: in the first case we fixed it at the estimated metallicity $Z = 0.2 Z_\odot$ for the SMC \citep{Russel&Dopita92}; in the second we left its value free to vary. We note that in both cases we obtained an equally good fit, which is slightly better than the one obtained with the PL+BB model (\chisqnu\ = 1.02 instead of 1.03); moreover, in the second case the best-fitting value of the abundance is 0.29$^{+0.22}_{-0.12}$, thus in agreement with the expected value for the SMC. In all cases the best-fitting absorption value is \nh\ $\simeq$ (1-2) $\times 10^{20}$ cm$^{-2}$, a value below the total Galactic absorption in the SMC direction estimated by \citet{DickeyLockman90} (\nh\ $\simeq 3.5 \times 10^{20}$ cm$^{-2}$). In fact, the fit of the same models with \nh\ fixed at this value resulted in a \chisqnu increase of with $\Delta$\chisqnu\ $\sim$ 0.1. Due to the high count statistics of the spectrum (d.o.f. $>$ 3000), this $\Delta$\chisqnu\ implies a significant worsening of the fit quality.

In the case of a PL+BB model, the BB normalization implies a radius of $\simeq$ 50 km for the emitting region, and a very small contribution ($\simeq$ 1.5 \%) to the total flux; the same contribution is higher ($\simeq$ 4.5 \%) in the case of the APEC model. The Gaussian components at $\simeq$ 0.5, 0.65, and 6.4 keV can be identified with N {\sc vii}, O {\sc viii}, and neutral Fe K$\alpha$ emission lines, respectively. For the N {\sc vii} line and (only in the case of the PL+BB model) the O {\sc viii} line the component width was consistent with 0, therefore we fixed this parameter. The Gaussian component at $\simeq$ 0.95 keV observed in the fit with the PL+BB model can be due to either a blend of several L$\alpha$ emission lines from Fe in a range of ionizations states (from {\sc Fe xviii} to Fe {\sc xx}) or a radiative recombination continuum (RRC) from O {\sc viii} - Ne {\sc ix}. The absorption component at $\simeq$ 1 keV can be attributed to Ne {\sc ix-x} or Fe {\sc xvi-xxiii}, while that at $\simeq$ 2.9 keV is consistent with a feature of S {\sc xiv-xv}.

\begin{table*}
\caption{Results of the simultaneous fit of the time-averaged spectrum of the \pn\ and MOS data. The double-component continuum consists of a power-law and either a blackbody or a thermal plasma model. In the second case both a fixed and a free metal abundance is considered. In addition, various Gaussian and GABS components are needed to account for, respectively, the positive and negative residuals in the spectrum.}\label{epic_fit}
\begin{center}
\begin{tabular}{cccc} \hline
Continuum Model							& \textsc{pl+bb}		& \textsc{pl+apec}			& \textsc{pl+apec}			\\
Parameter							& 				& (fixed abundance)			& (free abundance)			\\ \hline
\nh\ (10$^{20}$ cm$^{-2}$)					& 1.0$^{+0.1}_{-0.2}$		& 2.0$^{+0.2}_{-0.3}$			& 1.8$^{+0.4}_{-0.3}$			\\
$\Gamma$							& 0.87$^{+0.02}_{-0.01}$	& 0.83$^{+0.02}_{-0.01}$		& 0.83$\pm$0.02				\\
Flux$_{\rm PL}$ (0.2-12 keV, $\times 10^{-11}$\flux)		& 9.82$\pm$0.06 		& 9.77$^{+0.07}_{-0.06}$		& 9.77$\pm$0.03				\\
$kT_{\rm BB~or~APEC}$ (keV)					& 0.218$^{+0.013}_{-0.014}$	& 1.11$^{+0.12}_{-0.06}$		& 1.13$^{+0.10}_{-0.08}$		\\
$R_{\rm BB}$ (km) or $N_{\rm APEC}$ (cm$^{-5}$)			& 50$^{+6}_{-5}$		& 4.5$^{+0.8}_{-1.0} \times 10^{-3}$	& 4.0$^{+1.0}_{-1.2} \times 10^{-3}$	\\
Flux$_{\rm BB~or~APEC}$ (0.2-12 keV, $\times 10^{-12}$ \flux)	& 1.6$^{+0.2}_{-0.1}$		& 3.9$^{+0.7}_{-0.3}$			& 4.0$^{+0.3}_{-0.6}$			\\
Abundance (\textsc{apec})					& -				& 0.2 (fixed)				&	0.29$^{+0.22}_{-0.12}$		\\ \hline
\multicolumn{4}{c}{Emission lines}	\\
$E_{\rm line1}$ (keV)						& -				& 0.490$^{+0.015}_{-0.012}$		& 0.490$^{+0.008}_{-0.009}$		\\
$\sigma_{\rm line1}$ (keV)					& -				& 0 (fixed)				& 0 (fixed)				\\
Flux$_{\rm line1}$ ($\times 10^{-5}$ \fph)			& -				& 5.6$^{+3.5}_{-2.6}$			& 7.2$^{+3.5}_{-3.4}$			\\
EW$_{\rm line1}$ (eV)						& -				& 5.4$^{+3.5}_{-2.2}$			& 7.4$^{+4.1}_{-3.9}$			\\
$E_{\rm line2}$ (keV)						& 0.663$\pm$0.005		& 0.663$\pm$0.007			& 0.662$^{+0.007}_{-0.008}$		\\
$\sigma_{\rm line2}$ (keV)					& 0 (fixed)			& 0.020$^{+0.014}_{-0.012}$		& 0.025$^{+0.011}_{-0.016}$		\\
Flux$_{\rm line2}$ ($\times 10^{-4}$ \fph)			& 0.9$\pm$0.2			& 1.3$^{+0.3}_{-0.2}$			& 1.4$^{+0.5}_{-0.1}$			\\
EW$_{\rm line2}$ (eV)						& 11$^{+3}_{-4}$		& 17$^{+4}_{-5}$			& 19$^{+3}_{-4}$			\\
$E_{\rm line3}$ (keV)						& 0.957$^{+0.025}_{-0.024}$	& -					& -					\\
$\sigma_{\rm line3}$ (keV)					& 0.077$^{+0.021}_{-0.018}$	& -					& -					\\
Flux$_{\rm line3}$ ($\times 10^{-5}$ \fph)			& 8.1$^{+2.9}_{-2.8}$		& -					& -					\\
EW$_{\rm line3}$ (eV)						& 16$^{+6}_{-7}$		& -					& -					\\
$E_{\rm line4}$ (keV)						& 2.00$\pm0.07$			& 2.02$\pm$0.07				& 2.02$\pm$0.07				\\
$\sigma_{\rm line4}$ (keV)					& 0.25$^{+0.11}_{-0.07}$	& 0.24$^{+0.07}_{-0.06}$		& 0.24$^{+0.09}_{-0.06}$		\\
Flux$_{\rm line4}$ ($\times 10^{-4}$ \fph)			& 1.2$^{+0.5}_{-0.4}$		& 1.1$^{+0.5}_{-0.3}$			& 1.1$^{+0.4}_{-0.2}$			\\
EW$_{\rm line4}$ (eV)						& 52$^{+20}_{-15}$		& 48$^{+16}_{-19}$			& 47$^{+18}_{-15}$			\\
$E_{\rm line5}$ (keV)						& 6.43$^{+0.07}_{-0.06}$	& 6.42$^{+0.06}_{-0.05}$		& 6.42$\pm$0.06				\\
$\sigma_{\rm line5}$ (keV)					& 0.18$^{+0.10}_{-0.07}$	& 0.16$^{+0.07}_{-0.06}$		& 0.16$^{+0.07}_{-0.06}$		\\
Flux$_{\rm line5}$ ($\times 10^{-5}$ \fph)			& 4.9$^{+1.7}_{-1.4}$		& 4.3$^{+1.7}_{-1.2}$			& 4.3$^{+1.3}_{-1.2}$			\\
EW$_{\rm line5}$ (eV)						& 58$^{+16}_{-18}$		& 51$^{+17}_{-16}$			& 51$^{+14}_{-11}$			\\ \hline
\multicolumn{4}{c}{Absorption lines}	\\
$E_{\rm line6}$ (keV)						& -				& 0.99$^{+0.04}_{-0.01}$		& 1.00$\pm$0.03				\\
$\sigma_{\rm line6}$ (keV)					& -				& 0.08$^{+0.02}_{-0.03}$		& 0.08	$\pm$0.01			\\
Depth$_{\rm line6}$ ($\times 10^{-2}$ keV)			& -				& 3.4$^{+1.2}_{-1.9}$			& 4.7$^{+2.8}_{-2.0}$			\\
$E_{\rm line7}$ (keV)						& 2.89$^{+0.09}_{-0.04}$	& 2.88$^{+0.05}_{-0.04}$		& 2.88$^{+0.05}_{-0.04}$		\\
$\sigma_{\rm line7}$ (keV)					& 0.09$\pm$0.03			& 0.09$\pm$0.03				& 0.07	$\pm$0.03			\\
Depth$_{\rm line7}$ ($\times 10^{-2}$ keV)			& 2.1$^{+1.1}_{-0.8}$		& 1.5$^{+0.6}_{-0.5}$			& 1.5$\pm$0.06				\\ \hline
Flux$_{\rm BB~or~APEC}$/Flux$_{\rm PL}$ (0.01-12 keV)		& 1.6 \%			& 4.5 \%				& 4.5 \%				\\
Unabsorbed flux (0.2-12 keV, $\times 10^{-10}$ \flux)		& 1.001$^{+0.003}_{-0.004}$	& 1.022$^{+0.008}_{-0.007}$		& 1.022$\pm$0.009			\\
Luminosity (0.2-12 keV, $\times 10^{37}$ \lum)			& 3.55$\pm$0.01			& 3.63$^{+0.03}_{-0.02}$		& 3.63$\pm$0.03 			\\
\chisqnu/d.o.f.							& 1.03/3264			& 1.02/3261				& 1.02/3260				\\ \hline
\end{tabular}
\end{center}
\end{table*}

\section{RGS Spectroscopy}
\label{rgs}

We considered only the first-order spectra and ignored the second-order, since their count statistics was too limited for a meaningful spectral analysis. We extracted the first-order spectrum from the data of both RGS instruments. We verified that both spectra were consistent, then we combined them with the {\sc sas} task {\sc rgscombine}. The resulting spectrum was rebinned with a minimum of 30 counts per bin and analysed with {\sc xspec} in the energy range 0.33-2.2 keV.

\begin{figure}
\centering
\resizebox{\hsize}{!}{\includegraphics[angle=-90,clip=true]{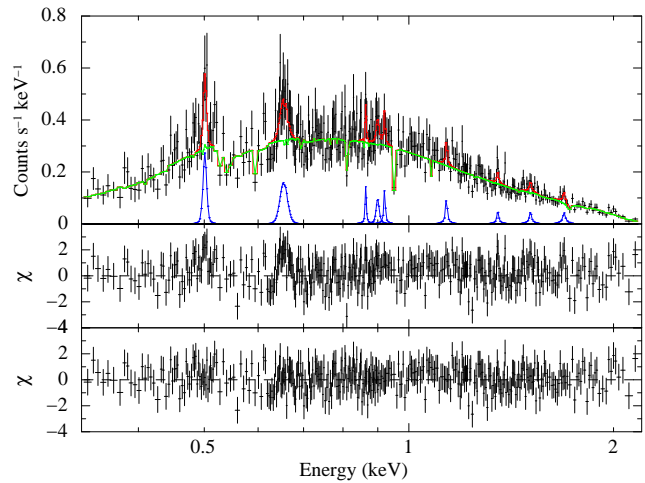}}
\caption{Combined RGS1 and RGS2 spectrum for the first order data. Upper panel: superposition of the spectrum with the best-fitting absorbed power-law model (green line) plus Gaussian components (blues lines, Table~\ref{RGS_parameters}). Middle panel: data-model residuals in the case of the fit with a simple power-law model. Lower panel: data-model residuals in the case of the best-fitting model.}
\label{RGS_spectrum}
\end{figure}

It was possible to describe the spectrum continuum with an absorbed power-law model, which however left several emission and absorption residuals. In fact, with this model we obtained \chisqnu/d.o.f. = 1.15/346. We modelled the emission features with Gaussian components and the absorption features with the {\sc gabs} model of {\sc xspec}. In this way, we found the main narrow emission features at $\simeq$ 0.5, 0.65, 0.86, 0.9, 0.92, 1.13, 1.35, 1.51, and 1.69 keV, and two absorption features at 0.81 and 1.72 keV. Their parameters are reported in Table~\ref{RGS_parameters}, while the RGS spectrum is shown in Fig.~\ref{RGS_spectrum}. In the case of the two absorption lines and of the emission lines at $\simeq$ 0.5, 0.65, and 0.9 keV, the line intrinsic width is larger than 0 and well determined. For the other lines the width is consistent with 0 or unconstrained, then in the spectral fit we fixed its value at 0. We verified that the normalizations of the absorption line at $\simeq$ 0.81 keV and of the emission lines at $\simeq$ 0.86, 0.92, 1.13 and 1.35 keV are significant at 90 \% confidence level only: therefore, we cannot claim a firm detection but only a hint for the presence of these lines. In Table~\ref{RGS_parameters} we report the most probable identification of each line. The lines at 0.5 and 0.65 keV were well constrained and can be associated with N {\sc vii} and O {\sc viii} Ly$\alpha$ lines, respectively. The two narrow lines at 0.9 and 0.92 keV are consistent with two components of the He-like Ne {\sc ix} emission triplet, while the line at 1.35 keV can be due to Mg {\sc xi}. The identification of the remaining emission and absorption features is more difficult, although all of them are consistent with various Fe ionization stages. For the best-fit final model, which includes all the emission and absorption features reported in Table~\ref{RGS_parameters}, we obtained \nh\ = (2$\pm$1)$\times 10^{20}$ cm$^{-2}$ and $\Gamma$ = 1.07$\pm$0.07, in agreement with the results obtained from the EPIC spectra. The corresponding \chisqnu/d.o.f. is 0.93/319, with a significant improvement compared with the simple absorbed power-law model.

For completeness, we also performed a fit of the RGS spectrum with the best-fitting models of the EPIC spectra described in section~\ref{EPIC}. The corresponding results are fully consistent with those previously shown, since the PL+BB or PL+APEC models used for the EPIC continuum can also describe the continuum component of the RGS spectra. However, we note that in both cases the RGS spectra show residuals, comparable to those reported in the middle panel of Fig.~\ref{RGS_spectrum}.

\begin{table*}
\caption{Best-fit parameters of the emission and absorption lines identified in the RGS spectrum of \IGR.}\label{RGS_parameters}
\begin{center}
\begin{tabular}{cccccc} \hline
Observed		& Ion						& Laboratory	& $\sigma$		& Flux$^{(a)}$ (10$^{-5}$ \fph)	& EW			\\
Energy			&						& Energy	& (eV)			& or				& (eV)			\\
(eV)			&						& (eV)		&			& Depth$^{(b)}$ (10$^{-2}$ keV)	&			\\ \hline
\multicolumn{6}{c}{Emission lines}																	\\
501$^{+1}_{-2}$		& N {\sc vii}					& 500.3		& 2.9$^{+2.8}_{-1.1}$	& 8.0$^{+3.1}_{-2.5}$		& 8.0$^{+2.6}_{-3.4}$	\\
654$^{+4}_{-3}$		& O {\sc viii}					& 653.5		& 7.7$^{+3.0}_{-2.3}$	& 8.3$^{+3.0}_{-2.2}$		& 11.1$^{+3.2}_{-3.9}$	\\
864$^{+2}_{-1}$		& Fe {\sc xviii-xxi} (?)			& -		& 0 (fixed)		& 1.8$\pm$1.2			& 3.0$\pm$1.9		\\
900$\pm$5		& Ne {\sc ix}					& 905.1		& 3.6$^{+9.5}_{-2.5}$	& 2.5$^{+1.5}_{-1.6}$		& 4.3$^{+2.9}_{-2.7}$	\\
921$^{+5}_{-4}$ 	& Ne {\sc ix}					& 922.1		& 0 (fixed)		& 2.0$^{+1.7}_{-1.6}$		& 3.7$\pm$2.9		\\
1135$^{+7}_{-4}$	& Fe {\sc xix-xxiii} (?)			& -		& 0 (fixed)		& 2.6$^{+2.2}_{-1.8}$		& 6.2$^{+4.1}_{-2.6}$	\\
1352$^{+4}_{-5}$	& Mg {\sc xi}					& 1352.0	& 0 (fixed)		& 2.1$^{+1.8}_{-1.6}$		& 6.0$^{+4.5}_{-4.9}$	\\
1509$\pm$7		& Fe {\sc xxi-xxiv} (?)				& -		& 0 (fixed)		& 2.9$^{+2.3}_{-1.5}$		& 9.6$^{+5.8}_{-6.5}$	\\
1691$^{+8}_{-6}$	& Fe {\sc xxii-xxiii} (?)			& -		& 0 (fixed)		& 4.5$^{+2.6}_{-2.1}$		& 18$^{+14}_{-13}$	\\ \hline
\multicolumn{6}{c}{Absorption lines}																	\\
811$\pm$2		& Fe {\sc xvii-xx} (?)				& -		& $<$ 1			& 2.3$\pm$2.1			& -			\\
1721$^{+15}_{-13}$	& Al {\sc xii-xiii} - Fe {\sc xxiii-xxiv} (?)	& -		& 5$^{+20}_{-4}$	& 3.7$\pm$3.1			& -			\\ \hline
\end{tabular}
\end{center}
$^{(a)}$ for the emission lines
\\
$^{(b)}$ for the absorption lines
\end{table*}

\section{Phase-resolved spectral analysis}
\label{resolved_spectral_analysis}

Since the folded light curves reported in Fig.~\ref{flc2E} and Fig.~\ref{flc4E} show that the spin profile of \IGR\ is strongly energy dependent, it is interesting to quantitatively explore the observed spectral evolution. For each EPIC camera we extracted one phase-selected spectrum in the phase ranges $\Delta\Phi$ = 0.2-0.5 (spectrum A) and another in $\Delta\Phi$ = 0.6-0.9 (spectrum B). They correspond, respectively, to the hardest (HR $>$ 0.9) and softest (HR $<$ 0.7) parts of the folded light curve.

Our aim was to assess whether these two spectra can be described with the same best-fitting model used for the time-averaged spectra and, in this case, if the best-fitting parameters have comparable or inconsistent values. Therefore, we performed an independent fit of the two spectra, adopting the PL+BB description of the spectral continuum. In both cases we fixed the absorption to the best-fit value obtained for the averaged spectrum (\nh\ = $1 \times 10^{20}$ cm$^{-2}$). The obtained results are reported in Table~\ref{2spectra}.

\begin{table}
\caption{Best-fit parameters of the EPIC spectra A and B, assuming the PL+BB description of the spectral continuum.}\label{2spectra}
\begin{center}
\begin{tabular}{cccc} \hline
Parameter					& Spectrum A			& Spectrum B			\\ \hline
\nh\ (10$^{20}$ cm$^{-2}$)			& 1 (fixed)			& 1 (fixed)			\\
$\Gamma$					& 0.83$\pm$0.01			& 0.90$\pm$0.02			\\
Flux$_{\rm PL}^{(a)}$				& 13.8$\pm$0.1 			& 7.79$^{+0.10}_{-0.09}$	\\
$kT_{\rm BB}$ (keV)				& -				& 0.31$\pm$0.01			\\
$R_{\rm BB}$ (km)				& -				& 40$^{+2}_{-3}$		\\
Flux$_{\rm BB}^{(b)}$				& -				& 4.8$^{+0.5}_{-0.3}$		\\ \hline
\multicolumn{3}{c}{Emission lines}		\\
$E_{\rm line1}$ (keV)				& -				& 0.49$\pm$0.01			\\
$\sigma_{\rm line1}$ (keV)			& -				& 0 (fixed)			\\
Flux$_{\rm line1}^{(c)}$			& -				& 1.3$^{+0.5}_{-0.2}$		\\
EW$_{\rm line1}$ (eV)				& -				& 13$^{+5}_{-4}$		\\
$E_{\rm line2}$ (keV)				& 0.664$^{+0.012}_{-0.014}$	& 0.654$^{+0.011}_{-0.012}$	\\
$\sigma_{\rm line2}$ (keV)			& 0 (fixed)			& $<$0.028			\\
Flux$_{\rm line2}^{(c)}$			& 1.0$^{+0.4}_{-0.3}$		& 1.3$^{+0.2}_{-0.3}$		\\
EW$_{\rm line2}$ (eV)				& 13$\pm$4			& 15$^{+6}_{-4}$		\\
$E_{\rm line3}$ (keV)				& 0.921$^{+0.046}_{-0.060}$	& -				\\
$\sigma_{\rm line3}$ (keV)			& 0.114$^{+0.061}_{-0.036}$	& -				\\
Flux$_{\rm line3}^{(c)}$			& 1.5$^{+0.7}_{-0.5}$		& -				\\
EW$_{\rm line3}$ (eV)				& 25$^{+8}_{-11}$		& -				\\
$E_{\rm line4}$ (keV)				& 2.01$^{+0.15}_{-0.17}$	& 2.07$\pm$0.07			\\
$\sigma_{\rm line4}$ (keV)			& 0.25 (fixed)			& 0.18$\pm$0.07			\\
Flux$_{\rm line4}^{(c)}$			& 1.5$\pm$0.5			& 1.3$\pm$0.5			\\
EW$_{\rm line4}$ (eV)				& 47$^{+19}_{-16}$		& 60$^{+21}_{-18}$		\\
$E_{\rm line5}$ (keV)				& 6.34$^{+0.06}_{-0.05}$	& 6.45$\pm$0.09			\\
$\sigma_{\rm line5}$ (keV)			& 0 (fixed)			& 0.24$^{+0.15}_{-0.10}$	\\
Flux$_{\rm line5}^{(d)}$			& 2.3$^{+1.4}_{-1.5}$		& 7.5$^{+3.2}_{-2.5}$		\\
EW$_{\rm line5}$ (eV)				& 19$^{+13}_{-12}$		& 115$^{+48}_{-35}$		\\ \hline
Flux$_{\rm BB}$/Flux$_{\rm PL}$ (0.01-12 keV)	& -				& 6.2 \%			\\
Unabsorbed flux$^{(e)}$				& 1.39$\pm$0.01			& 0.84$\pm$0.01			\\
Luminosity$^{(f)}$				& 4.93$\pm$0.04			& 2.98$\pm$0.03			\\
\chisqnu/d.o.f.					& 1.03/2889			& 1.03/2550			\\ \hline
\end{tabular}
\end{center}
$^{(a)}$ 0.2-12 keV, $\times 10^{-11}$ \flux
\\
$^{(b)}$ 0.2-12 keV, $\times 10^{-12}$ \flux
\\
$^{(c)}$ $\times 10^{-4}$ \fph
\\
$^{(d)}$ $\times 10^{-5}$ \fph
\\
$^{(e)}$ 0.2-12 keV, $\times 10^{-10}$ \flux
\\
$^{(f)}$ 0.2-12 keV, $\times 10^{37}$ \lum
\end{table}

In the case of spectrum A we found no evidence for the BB component. It is possible to describe the spectral continuum with a single absorbed power-law, while the emission features detected in the averaged spectrum can be observed also in this case. The source flux is almost completely due to the PL component. The four emission features are significant at 99 \% c.l. and their best-fitting energies and widths are consistent with those observed in the averaged spectrum. The only exception is the Fe line, which is characterized by a lower energy and width consistent with 0. Its equivalent width (EW) is much lower than in the averaged spectrum.

In the case of spectrum B we obtained a different result. First of all, here it is necessary to introduce a BB component, since a simple absorbed power-law can not describe the spectral continuum at soft energies. The flux of the PL component is much lower than in the previous case, while the contribution of the BB component to the total flux is larger than 6 \%. The BB temperature is signficantly higher than in the time-averaged spectrum. Moreover, in this spectrum we found no evidence of the emission feature at $\sim$ 0.9 keV. On the other hand, this spectrum shows an additional narrow emission feature at $\simeq$ 0.49 keV, which is undetected in the averaged spectrum: it can be described with a Gaussian component of intrinsic width fixed to 0, which is significant at 99 \% c.l. and can be attributed to N {\sc vii}. The properties of both the O and Fe lines are consistent with those observed in the averaged spectrum, but the EW of the Fe line is higher.

The independent fit of the two spectra shows that spectrum A is harder than spectrum B, although, in both cases, we obtained a value of the PL photon index consistent with that of the averaged spectrum. Moreover, neither in spectrum A nor in spectrum B we found any evidence of the absorption feature detected at 2.9 keV in the averaged spectrum. This is most probably due to the lower count statistics of the two spectra. The spectral variability between the two peaks can be ascribed to the variation of the relative contribution of the two continuum components, since the PL flux is higher in the first peak, while the BB flux is evident only in the second peak. In this way, it is possible to explain the energy-dependent pulse shape reported in Fig.~\ref{flc4E}. In addition, the strength of the Fe line increases in the second peak.

\begin{figure}
\centering
\resizebox{\hsize}{!}{\includegraphics[angle=-90,clip=true]{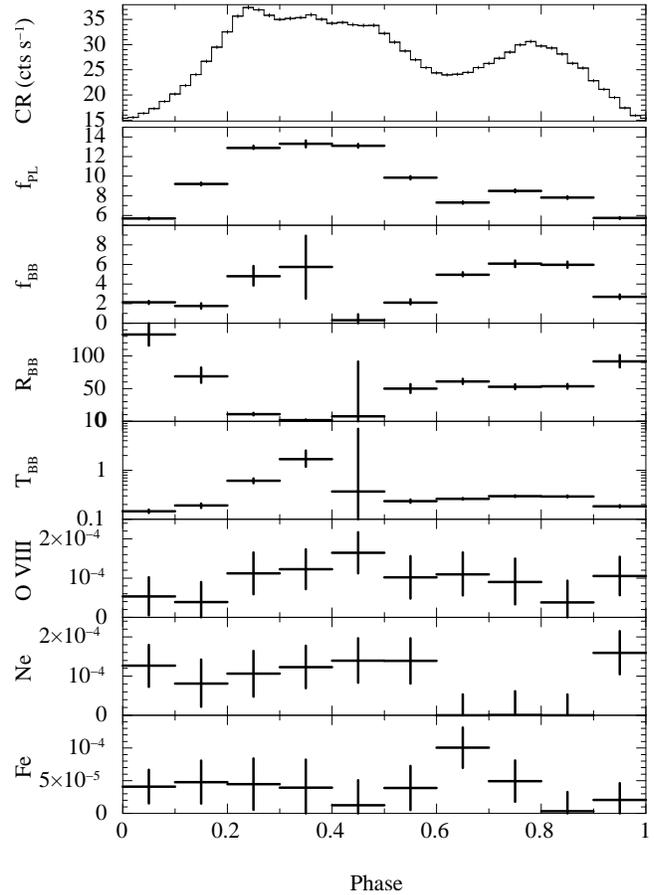}}
\caption{Variation with the pulse phase of the spectral components identified in the time-averaged spectrum of \IGR. From top to bottom, we report: 1) the source CR (in the energy range 0.2-12 keV); 2) the PL flux (in units of 10$^{-11}$ \flux); 3) the BB flux (in units of 10$^{-12}$ \flux); 4) the BB temperature (in keV); 5) the BB radius (in km); 6-8) the flux of the O, Ne, and Fe lines (in units of \fph).}
\label{resolved_spectra}
\end{figure}

To investigate in deeper detail the variability of the single spectral components, we divided the pulse profile in 10 equally spaced phase bins. Then, for each instrument we extracted 10 different spectra, one for each phase bin. We performed a simultaneous fit of the 10 sets of 3 spectra, assuming a common value for the absorption; moreover, we fixed the energy and width of the emission and absorption lines, since the count statistics is too low to constrain them in each phase bin. For the fit we considered three different solutions for the PL+BB continuum: (1) a common value of the PL spectral index and BB temperature, leaving both normalizations free to vary; (2) a common BB, leaving both the PL index and normalization free to vary; (3) a common value of only the PL index, leaving its normalization and both the BB parameters free to vary. We found that both solutions (1) and (2) are statistically unacceptable and can be dismissed, since in their case the Null Hypothesis Probability (NHP) is very low ($\sim 10^{-5} - 10^{-6}$). On the other hand, for solution (3) NHP $>$ 0.1. This result supports the hypothesis that the BB strongly varies along the pulse period. In Fig.~\ref{resolved_spectra} we report the fluxes of the spectral components as a function of the pulse phase. For completeness, we also report the BB radius and temperature. As expected, the profile of the PL flux reproduces that of Fig.~\ref{flc4E} at energies above 2 keV. On the other hand, the BB flux shows a clear peak in the phase range 0.6-0.9, while it is poorly constrained in the range 0.3-0.4 and consistent with 0 in the range 0.4-0.5. Between phases 0.3 and 0.5 the BB radius reaches its minimum values, while its temperature rapidly decreases. Also the emission lines show some hint of variability. The flux of the O {\sc viii} line gradually increases up to its maximum value in the phase range 0.4-0.5 (near the peak of the PL flux), then it decreases. The Ne {\sc ix-x} component is nearly constant, apart from the phase range 0.6-0.9, where its flux reduces to $\sim$ 0. Instead the Fe line has a different behaviour: it is almost constant over most of the pulse period, but has a marginal maximum in the phase range 0.6-0.7, at the beginning of the secondary flux peak of the continuum component.

 	 \section{\Swift/XRT Observations}
         \label{xrt}

\begin{figure}
\begin{center}
\includegraphics[width=8.5cm,angle=0]{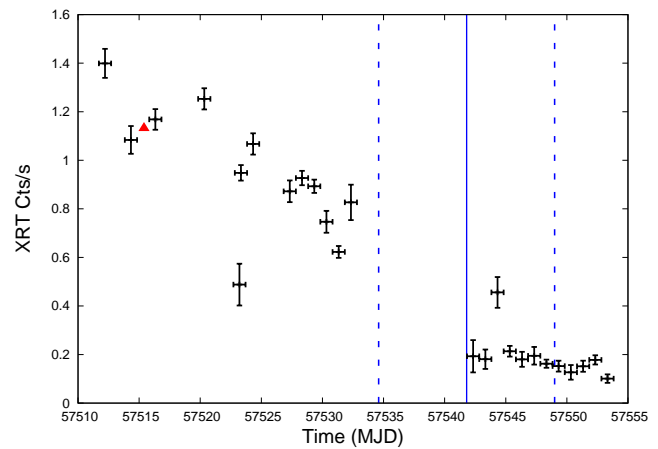}
\caption{Flux evolution of \IGR\ during the 2016 outburst, as observed with \Swift/XRT. The red triangle corresponds to the epoch and flux of the \XMM\ observation. The lines mark the estimated epoch of the centroid of the light-curve minimum, with its uncertainties.}
\label{swift}
\end{center}
\vspace{-0.75 cm}
\end{figure}

\Swift/XRT observed \IGR\ with 48 snapshot observations between 2016 May and June, with exposure times of a few hundred seconds up to $\sim$ 2.3 ks each. We analyzed the data by using the \Swift/XRT online tool\footnote{http://www.swift.ac.uk/user\_objects/} by \citet{Evans+09}. The long-term background-subtracted source lightcurve is shown in Fig.~\ref{swift}. The red triangle reports the \Swift/XRT count rate corresponding to the source flux measured with \XMM. For reference, in the figure we report as a solid line at MJD = 57541 the estimated epoch of the centroid of the light-curve minimum, which is based on the ephemerides provided by \citet{Segreto+13} and the orbital period of 35.1 d. The two dashed lines are the estimated lower and upper limits for the epoch of the centroid, taking into account all the uncertainties. The peak of the outburst was reached at MJD $\sim$ 57512. Then the source flux decayed in $\sim$ 30 days reaching a constant flux value, lower than the peak flux by a factor of 7. We created two stacked spectra for the high and low flux intervals that we selected as the epochs before and after MJD = 57540, respectively. Both spectra can be well fitted with a single absorbed power law. Since the column density was unconstrained, we fixed it to 3.5$\times10^{20}$ cm$^{-2}$ and obtained a photon index $\Gamma_{\rm high} = 0.9 \pm 0.1$ and $\Gamma_{\rm low} = 0.71 \pm 0.08$ for the high and low flux spectra, respectively. This suggests a very marginal spectral variability.

	      \section{Discussion}\label{sec:discussion}

The new outburst of \IGR\ in April 2016 was the first after that observed with INTEGRAL and \Swift\ in 2008, in which the source was discovered. While the 3-10 keV flux measured with INTEGRAL was $f_{\rm 2008} = 1.6 \times 10^{-12}$ \flux, the flux in the same energy range measured with \XMM\ was $f_{\rm 2011} = 6 \times 10^{-11}$ \flux, i.e. $\simeq$ 40 times higher. Therefore, thanks to the large collecting area and high spectral resolution of \XMM, our ToO observation allowed us to investigate at an unprecedented level of detail this poorly studied source. While the previous observations were limited at energies above 3 keV, we extended both the timing and spectral analyses down to $\simeq$ 0.2 keV. In Table~\ref{transients} we report the main properties of this source together with those of \rxj\ \citep{Sidoli+15} and \smc\ \citep{LaPalombara+16}, since they are the other two transient Be binary pulsars in the SMC that we have observed at high spectral resolution during an outburst. Compared to \IGR, these two sources have a shorter pulse period ($\simeq$ 2.5 s instead of $\simeq$ 11.5 s) and were observed at a higher luminosity level (a factor $\simeq$ 2 and $\simeq$ 4 for \rxj\ and \smc, respectively). However, their spectral and timing properties are very similar to those observed in \IGR\ and, hence, it is interesting to directly compare these three sources.

\begin{table}
\caption{Comparison of the main timing and spectral parameters observed in the transient BeXRBs \rxj\ \citep{Sidoli+15}, \smc\ \citep{LaPalombara+16}, and \IGR\ (this work).}\label{transients}
\begin{center}
\begin{tabular}{cccc} \hline
Parameter				& \rxj\				& \smc\				& \IGR\	\\ \hline
$L_{\rm X}^{(a)}$			& 7				& 14				& 3.6	\\
$P_{\rm spin}$ (s)			& 2.76				& 2.37				& 11.58	\\
$\pdot_{\rm spin}^{(b)}$		& -1.27				& 0.66				& 17.16	\\
PF (\%)					& 8.9				& 35				& 43	\\
\nh\ (10$^{20}$ cm$^{-2}$)		& 2.3$^{+0.6}_{-0.5}$		& 18$\pm$3			& 1.0$^{+0.1}_{-0.2}$			\\
$kT_{\rm BB}$ (eV)			& 93$\pm$5			& 135$^{+14}_{-11}$		& 218$^{+13}_{-14}$	\\
$R_{\rm BB}$ (km)			& 350$^{+80}_{-50}$		& 320$^{+125}_{-95}$		& 50$^{+6}_{-5}$	\\
f$_{\rm BB}$/f$_{\rm PL}$ (\%)		& 1.7				& 3.1				& 1.6	\\
$kT_{\rm APEC}$ (keV)			& 0.21$\pm$0.03			& 1.22$^{+0.07}_{-0.10}$	& 1.13$^{+0.10}_{-0.08}$	\\
$N_{\rm APEC}^{(c)}$			& 25$^{+8}_{-6}$		& 5$\pm$1			& 4$\pm$1	\\
f$_{\rm APEC}$/f$_{\rm PL}$ (\%)	& 7				& 1.8				& 4.5	\\
N {\sc vii}				& yes				& yes				& yes	\\
O {\sc vii}				& no				& yes				& no	\\
O {\sc viii}				& yes				& yes				& yes	\\
Ne {\sc ix}				& yes				& yes				& yes	\\
Ne {\sc x}				& no				& yes				& no	\\
Mg {\sc xi}				& no				& no				& yes	\\
Si {\sc xiii}				& no				& yes				& no	\\
Si {\sc xiv}				& no				& yes				& no	\\
E$_{\rm Fe-K}$ (keV)			& 6.6				& 6.6				& 6.4	\\
$d_{\rm BB}$ (km)			& 3000				& 1800				& 400	\\
$R_{\rm m}$ (km)			& 900				& 740				& 1100	\\ \hline
\end{tabular}
\end{center}
$^{(a)}$ 0.2-12 keV, $\times 10^{37}$ \lum
\\
$^{(b)}$ $\times 10^{-12}$ s s$^{-1}$
\\
$^{(c)}$ $\times 10^{-3}$ cm$^{-5}$
\end{table}


The spin period of \IGR\ measured with \XMM\ was $P_{\rm spin, 2016}$ = 11.58208(2) s. Compared to the period obtained with \RXTE\ at the epoch of the previous outburst ($P_{\rm spin, 2008}$ = 11.57809(2) s, \citealt{McBride+10}), which was measured on MJD = 54824, this period implies a difference $\Delta P_{\rm spin} \simeq$ 4 ms. It is possible that this difference is due to the orbital motion of the NS around the Be star, since the photon arrival times were not corrected for this effect. We note that the \XMM\ and \RXTE\ observations were performed at two different orbital phases: in fact, taking into account the uncertainties on the date of the reference orbital phase (0.7 days) and on the orbital period (0.1 days), we estimated $\Phi_{\rm orbit, XMM} = 0.25 \pm 0.21$ and $\Phi_{\rm orbit, RXTE} = 0.57 \pm 0.05$. Assuming a mass $M \sim 10 \msole$ for the Be star, the orbital period $P_{\rm orb} \simeq$ 35 d implies an average orbital velocity $v_{\rm orb} \simeq$ 140 km/s, that can account for the spin-period difference between the two observations.


The source pulsations were clearly seen over the whole energy range, down to E $<$ 0.5 keV, and the pulse profile shows a double peak in each energy range (Fig.~\ref{flc2E}). This is at odds with the results obtained with \RXTE\ in 2008, which detected only a single, broad peak \citep{McBride+10}, but agrees with what we observed for \rxj\ and \smc, which show both a double-peaked profile. The pulse profile observed with \XMM\ is strongly energy dependent, since the prominence of the second peak decreases with energy (Fig.~\ref{flc4E}): it dominates the first peak at E $<$ 0.5 keV but is very low at E $>$ 4 keV. We note that we obtained a similar result for \rxj, where the second peak emerges only below 4 keV \citep{Sidoli+15}. Finally, in \IGR\ also the pulsed fraction depends on the energy range, since it increases up to $\sim$ 50 \% above 2 keV. A similar high pulsed fraction was also observed in the \RXTE\ observation of 2008 \citep{McBride+10}. This is a common characteristics of transient pulsars during their outbursts, since similar values were also noticed in \rxj\ \citep{Kohno+00} and in \smc\ \citep{Corbet+01,Yokogawa+01,LaPalombara+16}.

The average EPIC spectrum of \IGR\ could be roughly fitted with a rather hard ($\Gamma \simeq$ 0.8-0.9) power-law component, which however is softer than the cut-off PL component ($\Gamma$ = 0.4$\pm$0.2) used to describe the combined \Swift/XRT and IBIS spectrum taken in 2008 \citep{McBride+10}. Although the PL component dominates the source spectrum over the whole energy range, below $\sim$ 1 keV an additional low-energy component was necessary to fit the observed data excess. We described this soft excess either with a soft (kT $\sim$ 0.2 keV) blackbody or with a hot (kT $\sim$ 1 keV) thermal plasma model, which contributes for only a few per cent to the total source luminosity. Therefore, the properties of this component are very similar to those of the soft component observed in \rxj\ and \smc\ (Table~\ref{transients}), although the luminosity of \IGR\ is lower than that of the other two sources. This luminosity difference, and the higher best-fit value of $T_{\rm BB}$, are the main reasons for the lower value of $R_{\rm BB}$ compared to the other two sources. As in the case of \rxj, also for \IGR\ the best-fit \nh\ value is lower than the estimated absorption value in the SMC direction. For both sources fixing \nh\ = 3.5$\times10^{20}$ cm$^{-2}$ \citep{DickeyLockman90} resulted in a worse fit. A difference of a factor of a few between the equivalent column density derived from X-ray spectral fits and that estimated from the neutral hydrogen measurements is not uncommon. Besides the systematic uncertainties in the two measurements, one has to consider the structure of the interstellar medium on small angular scales and the fact that X-ray absorption and 21 cm line are caused by different components of the ISM.


The analysis of the RGS spectrum revealed the presence of narrow emission lines, due to N, O, Ne, and Mg, and of some other emission and absorption features, probably due to L-shell from Fe at various ionization levels. Moreover, the EPIC spectrum showed a clear emission line at $\simeq$ 6.4 keV, which has an EW $\sim$ 50 eV and can be attributed to K-$\alpha$ emission from neutral iron. While most of the narrow RGS features were also observed in \rxj\ and \smc\ (Table~\ref{transients}), this is not true for the EPIC iron line: in fact, in both these sources the energy of the detected feature was $\simeq$ 6.6 keV, thus consistent with emission from ionized Fe. We note that an emission feature consistent with K-$\alpha$ line from neutral Fe was observed in 2000 in \smc, when \ASCA\ detected this source at a low luminosity level ($L_{\rm X} \sim 4 \times 10^{36}$ \lum).

We performed a phase-resolved spectral analysis of the EPIC data in order to investigate the energy dependence of the pulse profile and the observed spectral variability with the pulse phase. We found that the source flux in the first peak is entirely due to the PL component, since there is no evidence of an additional soft component. On the other end, this component is clearly present in the second peak, where the PL flux is signficantly lower than in the first peak (Table~\ref{2spectra}). Fig.~\ref{resolved_spectra} shows that the soft component varies in phase with the soft light curve. This soft component has a smooth pulse shape, with a peak at $\Phi \simeq$ 0.75: it is shifted of $\Delta\Phi \sim$ 0.5 from the first peak (the one due to the hard component), and its width ($\Delta\Phi \sim$ 0.3) is a small fraction of the whole pulse period. This properties suggest that the two components are related but are due to different emission processes, which occur in different regions around the central NS. Also the other spectral features show some variability: both the O and Fe lines are detected in both the pulse peaks, but the EW of the Fe line is much higher in the second, where the soft component reaches its maximum flux; moreover, the broad component at $\sim$ 0.95 keV is detected only in the first peak, while the N line is detected only in the second peak. These results suggest that the variability of the N and Fe lines is linked to that of the soft component.


Since the estimated luminosity of \IGR\ is $L_{\rm X} \simeq 3.5 \times 10^{37}$ \lum, according to \citet{Hickox+04} it is possible that a significant fraction of the observed soft excess is due to thermal emission from a hot plasma. This hypothesis is partially supported by the results of the EPIC spectral analysis, since the PL+APEC model provides a good fit of the spectral continuum; moreover, if the model metallicity is left free to vary, its best-fit value ($Z = 0.29^{+0.22}_{-0.12} Z_\odot$) is consistent with the estimated metallicity of the SMC ($Z = 0.2 Z_\odot$). However, the normalization of the APEC component ($\simeq$ 0.004) implies an emission measure $n^2 V \simeq 1.4 \times 10^{59}$ cm$^{-3}$. Since the expected gas density should be $n < 10^{12}$ cm$^{-3}$, the radius of the emitting spherical region for the optically thin plasma should be $R \gsim 3.2 \times 10^{11}$ cm, corresponding to a light travel time of $\simeq$ 11 s. Although this value is very close to the pulse period, it is much higher than the variability timescale of the soft component (Fig.~\ref{flc4E} and \ref{resolved_spectra}); hence, it would be very difficult to explain the variability properties previously described with emission from a diffuse plasma. Moreover, for both the EPIC and RGS spectra the fit of the spectral continuum with the PL+APEC model leaves several residuals, which correspond to spectral lines of various elements. Therefore, although we cannot exclude the presence of a hot thermal plasma, it is difficult that it is the origin of the soft excess. For this reason, in the following we consider the alternative possibility of reprocessing of the primary emission in a optically thick region.


The high accretion rate on to the pulsar during the source outburst can very likely cause the formation of an accretion disk around the NS. In this case, the inner edge of the disk would be a natural reprocessing site for the primary X-rays emitted from the accreting pulsar \citep{Hickox+04}. The ratio between the reprocessed and the primary luminosity would depend on $\Omega$, the solid angle subtended by the reprocessing region: $L_{\rm BB}$ = ($\Omega$/4$\pi$) $L_{\rm X}$. On the other hand, if $d_{\rm BB}$ is the distance of the reprocessing site from the source of the primary emission, the BB luminosity is given by the relation $L_{\rm BB} = \Omega d_{\rm BB}^2 \sigma T^4_{\rm BB}$. Therefore, the distance $d_{\rm BB}$ can be estimated from the relation $d_{\rm BB}^2 = L_{\rm X}$/($4\pi \sigma T^4_{\rm BB}$). In the case of \IGR\ $L_{\rm X} \simeq 3.5 \times 10^{37}$ \lum and $T_{\rm BB} \simeq$ 0.2 keV, which imply $d_{\rm BB} \simeq 4 \times 10^7$ cm. The inner edge radius should be comparable to the magnetospheric radius $R_{\rm m} \sim 1.5 \times 10^8 m^{1/7} R_6^{10/7} L_{37}^{-2/7} B_{12}^{4/7}$ cm, where $m$ is the NS mass in units of solar masses, $R_6$ is the NS radius in units of $10^6$ cm, $L_{37}$ is the X-ray luminosity in units of $10^{37}$ \lum, and $B_{12}$ is the NS magnetic field in units $10^{12}$ G \citep{Davies&Pringle81}. Assuming $m$ = 1.4, $R_6$ = 1 and $B_{12}$ = 1, for \IGR\ we obtain $R_{\rm m} \simeq$ 1100 km, a value which is comparable to those obtained for \rxj\ and \smc\ (Table~\ref{transients}). The differences among the estimated values of $R_{\rm m}$ for these sources are due essentially to the different luminosity levels, since in all cases we assume the same values for the NS mass, radius and magnetic field. Therefore, the magnetospheric radius decreases as the luminosity increases. We note that, for all these source, $d_{\rm BB}$ is comparable with $R_{\rm m}$ within a factor 2--3. This discrepancy can be due to geometrical factors, since a tilted and/or warped accretion disk can affect our estimate of $R_{\rm m}$; moreover, the NS magnetic field can be different from $10^{12}$ G.


Based on the above described timing and spectral properties of \IGR, we attribute both the soft excess and the iron line to the reprocessing of the primary X-rays from the inner edge of the disc. Since the primary emission is beamed, along the pulse period it sweeps the inner edge of the disc illuminating only a limited section of its surface in each moment; moreover, due to the disk geometry, only a limited fraction of the disk edge is visible for us. Therefore, we can see the reprocessed component only when the primary beam hits this visible portion of the disk. On the other hand, the narrow lines observed in the RGS spectrum are not pulsed and are most probably due to photoionized plasma in regions above the disc.

\section*{Acknowledgments}
We acknowledge financial contribution from the agreement ASI-INAF I/037/12/0. NLP and LS acknowledge the grant from PRIN-INAF 2014 `Towards a unified picture of accretion in HMXRBs'. PE acknowledges funding in the framework of NWO Vidi award A.2320.0076.

\bibliographystyle{mn2e} 
\bibliographystyle{mnras}
\bibliography{biblio}

\bsp

\label{lastpage}

\end{document}